\documentclass{ifacconf}

\usepackage{graphicx}      % include this line if your document contains figures
\usepackage{natbib}        % required for bibliography
\usepackage{amsfonts}
\usepackage{amsmath}
\usepackage{amssymb}
\usepackage{mathtools}
\usepackage{color}
\usepackage{comment}
\usepackage{booktabs}
\usepackage{epstopdf}
\usepackage{colortbl}
\usepackage{url}
\usepackage{todonotes}

\begin{document}
\begin{frontmatter}

\title{Continuous Time-Delay Estimation From Sampled Measurements} 
%\thanksref{footnoteinfo}

\thanks[footnoteinfo]{A. Medvedev was partially supported the Swedish Research Council under grant 2019-04451.}

\author{Mohamed  Abdalmoaty and Alexander Medvedev}
\address{Department of Information Technology, Uppsala University, PO Box 337, SE-75105, Uppsala, Sweden. (e-mails:\{mohamed.abdalmoaty,alexander.medvedev\}@it.uu.se)}

%\address[First]{National Institute of Standards and Technology, 
%   Boulder, CO 80305 USA (e-mail: author@ boulder.nist.gov).}
%\address[Second]{Colorado State University, 
%   Fort Collins, CO 80523 USA (e-mail: author@lamar. colostate.edu)}
%\address[Third]{Electrical Engineering Department, 
%   Seoul National University, Seoul, Korea, (e-mail: author@snu.ac.kr)}

\begin{abstract}                % Abstract of not more than 250 words.
An algorithm for continuous time-delay estimation from sampled output  data and known input of finite energy is presented.  The continuous time-delay modeling allows for the estimation of subsample delays. The proposed estimation algorithm consists of  two steps. First, the continuous Laguerre spectrum of the output signal is estimated from discrete-time (sampled) noisy measurements. Second, an estimate of the delay value is obtained in Laguerre domain given a continuous-time description of the input. The second step of the algorithm is shown to be intrinsically biased, the bias sources are established, and the bias itself is modeled. The proposed delay estimation approach is compared in a Monte-Carlo simulation with state-of-the-art methods implemented in time, frequency, and  Laguerre  domain demonstrating comparable or higher accuracy for the considered case.
\end{abstract}

\begin{keyword}
Delay, continuous systems, sampled data, estimation, bias
\end{keyword}

\end{frontmatter}
%===============================================================================

\section{Introduction}
The problem of (pure) time-delay estimation has been repeatedly addressed over the years but still cannot be seen as solved completely. It has numerous application areas that require different properties of the estimator regarding its accuracy, computational complexity, robustness to a certain class of disturbances and channel attenuation. 

There are also significant differences between time-delay estimation methods intended for controller design \citep{Fridman:2004} and signal processing applications. In modeling for control, an overestimate of the delay is preferred to an underestimate due to concerns regarding closed-loop stability. Besides, by enforcing controller robustness, the sensitivity of the closed-loop system to delay uncertainty can  be effectively minimized. In signal processing, time-delay estimation (time-of-arrival) constitutes the basis of imaging and distance measurement technology. Radars, sonars, and lidars \citep{M92}, as well as ultrasound applications \citep{Svilainis2019}, very much rely on the accuracy of the time-delay estimator, and the sampling rate of the involved signals is a clear bottleneck in enhancing performance. Measuring the time-delay in sampling time is therefore often insufficient and estimating subsample delays becomes a necessity. 

Several approaches to time-delay estimation have been covered in the literature. Perhaps the most common ones are the generalized cross-correlation approach (see e.g., \cite{Carter1987, Moddemeijer1991, Jacovitti1993}) and the two-step parameter estimation approach (see e.g., \citet{Chan1981, Reed1981, So2001, Wen2015}). The generalized cross-correlation method is a conventional method where the estimator is given by the maximizer of the cross-correlation function between the reference signal and its delayed version. To obtain subsample delay estimates, interpolation in time or frequency domain is used \citep{Cespedes1995,pinton2006,viola2008, Svilainis2019}. The two-step parameter estimation method (a.k.a. maximum-likelihood approach to interpolation) is based on estimating a non-causal FIR model  of the delay operator in time domain, and then using sinc interpolation to interpolate the tap coefficients and derive a subsample estimate; see e.g., \cite{Wen2015}. 

Orthonormal functional bases, e.g. Laguerre and Kautz, are extensively used in systems theory for parsimonious representation of dynamics and signals \citep{HHW05}.  The goal of their use is generally to approximate infinite-dimensional dynamics with finite-dimensional these; see e.g. \cite{MP99}. The concept of (continuous) time-delay estimation based on Laguerre spectra of the input and output has been introduced in \cite{FM99}. The estimation algorithm presented there is essentially least-squares exploiting the shift structure of the regressor in Laguerre domain. 
A time-delay estimation approach combining Pad{\' e} approximation of the delay operator with a Laguerre basis representation of the model is proposed in \cite{IHD01}. In a benchmark comparison of delay estimation methods reported in \cite{BL03}, the algorithms based on Laguerre functions were commended for their robustness. 

This paper continues along the path initiated in \cite{Abdalmoaty2022} where the problem of estimating a discrete delay, in a stochastic framework, based on Laguerre spectra of the input and output was addressed.  The main contributions are as follows: i) An algorithm for estimating the Laguerre spectrum of a continuous signal from its sampled measurements is proposed ii) A time-delay estimator making use of the estimated Laguerre spectrum of the output is devised. It is shown to be intrinsically biased and the sources of this bias are identified and analysed iii) A Monte-Carlo numerical performance analysis of the proposed algorithm suggests comparable or higher accuracy compared to alternative state-of-the-art approaches. 
Despite a clearly demonstrated potential, the proposed method is not meant to replace existing ones, but rather offer an alternative  that may be suitable in some applications.

The rest of the paper is organized as follows. First, background necessary for the exposition  on continuous Laguerre functions is provided. Then, the delay estimation problem is formulated in a stochastic framework. Further, the proposed method is explained in detail and related to some available approaches in the literature. An extensive simulation study illustrates the accuracy of the delay estimate and the impact  of the user-selected parameters in the algorithm. Finally, conclusions are drawn. 
 
\section{Background}\label{sec:background}
\subsection{Laguerre domain}
The Laplace transform of the $k$-th  continuous Laguerre function  is given by
\begin{equation}\label{eq:basis}
    \ell_k(s)=\frac{\sqrt{2p}}{s+p}T^k(s), \quad T(s)\triangleq\frac{s-p}{s+p},
\end{equation}
for $s \in \mathbb{C}$, $k \in \mathbb{N}$, where $p>0$ represents the Laguerre parameter, and $T(s)$ is the continuous Laguerre shift operator. 

Let $\mathbb{H}_c^2$  be the Hardy space of  functions analytic in the open left half-plane.
The set $\{\ell_k \}_{k \in \mathbb{N}}$ is an orthonormal complete basis in $\mathbb{H}_c^2$  with respect to the inner product 
\begin{align}\label{eq:inner}
\langle W, V \rangle \triangleq \frac{1}{2\pi i}\int_{-\infty}^{\infty}W(s)V(-s)~{\rm d}s.
\end{align}

Any function $W \in \mathbb{H}^2_c$  can be represented as a Laguerre series
\begin{equation}\label{eq:spectrum}
    W(s)=\sum_{k=0}^\infty w_k \ell_k(s),  \quad  w_j=\langle W, \ell_j \rangle,
\end{equation}
and the set $\{w_j\}_{j\in\mathbb{N}}$ is then  referred to as the {\em continuous Laguerre spectrum} of $W$, or simply the Laguerre spectrum of $W$.

\subsection{Time-delay  in Laguerre domain}\label{sec:time_delay}
The well-known associated Laguerre polynomials  (see e.g.  \cite{S39})   are explicitly given by
\begin{equation}\label{eq:assoc_Laguerre}
\mathrm{L}_m(\xi; \alpha)=\sum_{n=0}^m \frac{1}{n!} \binom{m+\alpha}{m-n} (-\xi)^n, \ \forall m\in \mathbb{N}, \quad \xi\in\mathbb{R}.
\end{equation}
In what follows, only the polynomials with a particular value of $\alpha$ are utilized and the shorthand notation $\mathrm{L}_m(\xi)\triangleq\mathrm{L}_m(\xi; \alpha)|_{\alpha=-1}$ is introduced. 
The associated Laguerre polynomials obey the following three-term relationship \citep{R60}
\begin{equation}\label{eq:three-term-cont}
\mathrm{L}_{m+1}(\xi)= \frac{1}{m+1}(\xi+2m)\mathrm{L}_m(\xi) - \frac{m-1}{m+1} \mathrm{L}_{m-1}(\xi).
\end{equation}

Consider the signal $u(t) \in \mathbb{L}_2\lbrack 0, \infty)$ given by its Laguerre spectrum $\{ u_j \}_{j\in\mathbb{N}_0}$.  Being passed through a pure delay block 
\begin{equation}\label{eq:cdelay}
y(t)=u(t-\tau), \quad \tau\ge 0, \quad t\in \lbrack 0, \infty),
\end{equation}
 the input $u(t)$ results in the output  $y(t) \in \mathbb{L}_2\lbrack 0, \infty)$ with the spectrum $\{ y_j \}_{j\in\mathbb{N}_0}$. Then, according to \cite{Hidayat12}, the following relationship holds  between the spectra
\begin{align}\label{eq:poly_pure_delay}
y_j = \sum_{k=0}^{j-1} h_{j-k}(\kappa) u_{k}+ h_0(\kappa) u_j , \ \forall j \in \mathbb{N}_0,  
\end{align}
where $h_k(\kappa)=\mathrm{e}^{-\frac{\kappa}{2}}\mathrm{L}_k(\kappa)$ and  $ \kappa=2p\tau$.
Notice that $\mathrm{L}_0(\kappa)=1$ and, therefore,  $h_0(\kappa)=\mathrm{e}^{-\frac{\kappa}{2}}$. The polynomials $h_i, i=0,1,\dots$, by an analogy with linear time-invariant systems, can be interpreted as the Laguerre-domain Markov parameters of  delay operator \eqref{eq:cdelay}. Since the operator is infinite dimensional, a Hankel matrix of the Markov parameters always has full rank.

\section{Delay estimation problem}
Consider a case where  the output signal measurements of \eqref{eq:cdelay} are sampled with the sampling time  $\Delta>0$  and corrupted by noise.  The observations at the discrete-time instances $t_n=n\Delta$,  $n=0,1,\dots, N-1$ are thus given by 
\begin{equation}\label{eq:measurement_model}
\begin{aligned}
z_n&=y(t_n)+e_n,\\
& = u(t_n -\tau) + e_n,
\end{aligned}
\end{equation}
where $e_n$ is a discrete-time measurement noise with zero mean and known variance $\lambda$. Further assumptions on the noise will be introduced later.  

 Given the data triplet 
\[
D_N = \left( \left\{z_n\right\}_{n=0}^{N-1} ,   \left\{u_k\right\}_{j=0}^I, p \right),
\]
where $\{u_j\}_{j=0}^I$, for some finite positive integer $I$, is the finite continuous Laguerre spectrum of the input with respect to a fixed Laguerre parameter value $p$ in \eqref{eq:basis}, the problem considered in this paper is to devise an estimator of the delay $\tau\in \mathbb{R}_+$: $D_N \mapsto \hat{\tau}  $.

Assume that $u(t)$ is non-zero on the time interval $[0, T_u]$.  Let $\tau_\text{max}$ be an upper bound on the unknown delay to be estimated. Then the observation/measurement interval is chosen as $[0,T]$ with $T = T_u + \tau_\text{max}$.

The Laguerre spectrum of $y$  defined by \eqref{eq:spectrum} can, in time domain, be evaluated by calculating the projections
\begin{equation}\label{eq:spectrum_definition}
y_j= \int_{0}^{\infty} y(t) \ell_j(t)~\mathrm{d}t.
\end{equation}
Since the continuous function $y(t)$ is not available, it has to be reconstructed from the discrete-time noisy measurements $\{z_n\}_{n=0}^{N-1}$. A straightforward  way of doing this is to apply (polynomial) interpolation to the measured sequence and then numerically evaluate integral \eqref{eq:spectrum_definition}. This approach disregards the presence of noise, and  generally leads to biased estimates of $y_j$.

Another approach is to assume a continuous signal model of the output as a linear combination of the Laguerre functions (can be written as an LTI system), and estimate the Laguerre coefficients by implementing a sampled-data algorithm. Then a stochastic noise model can be exploited to minimize the estimate variance.

\section{Maximum-Likelihood, frequency interpolation, and the CRLB}\label{sec:MLE}
\subsection{Maximum-Likelihood Estimation}
\subsubsection{Laguerre domain:}
Suppose that the measurement noise is Gaussian with a constant variance $\lambda$. Then the measurements $z_n$ are a realization of a Gaussian process with a mean function
\[
\begin{aligned}
\mu_n(\tau)  &= u(t_n -\tau) = y(t_n) = \sum_{j=0}^\infty y_j \ell_j(t_n)\\
&= \sum_{j=0}^\infty \;\ell_j(t_n)\sum_{k=0}^{j-1} \left(h_{j-k}({\kappa}) u_{k}+ h_0({\kappa})u_j\right).  \\
\end{aligned}
\]
The negative log-likelihood function of $\tau$ is such that
\[
- \log p (Z ;\tau)  \propto \Delta \sum_{n=0}^{N-1} (z_n - \mu_n(\tau) )^2,
\]
where $Z \triangleq \begin{bmatrix}z_0& \dots & z_{N-1} \end{bmatrix}^\top$, and hence the Maximum-Likelihood (ML) estimator of $\tau$ is defined as the minimizer of the mean square error between the observations and the mean function.  

Recall that $\kappa = 2 p \tau$. Then, from the definition of $h_k(\cdot)$ and \eqref{eq:three-term-cont}, one has
\[
\begin{aligned}
\partial_\tau h_k(\tau) &= -p e^{-p \tau} L_k (\tau) + e^{-p\tau} \partial_\tau L_k(\tau),\\
\partial_\tau L_k(\tau) &= \frac{2((k-1)p\tau) - k}{m}\partial_\tau L_{k-1}(\tau) -\frac{2p}{k} L_{k-1}(\tau)  \\
& \qquad + \frac{k-2}{k} \partial_\tau L_{k-2}(\tau);
\end{aligned}
\]
these expressions can be utilized to evaluate the gradient of the negative log-likelihood function.

\subsubsection{Time domain:}
Because the input signal is assumed to be constructed in  Laguerre domain and possess a finite Laguerre spectrum by design, it holds that
\[
{u}(t) = \sum_{k=0}^I u_j {\ell}_j(t)
\]
and, consequently, the mean function of the observations and its gradient $\partial_\tau \mu_n(\tau)$ can be evaluated in closed form. It holds that
\[
\partial_\tau u(t_n-\tau) = \sum_{k=0}^I u_j  \partial{\ell}_j(t_n-\tau),
\]
where $\partial{\ell}_j(t_n-\tau) = 0 \; \forall j \in \mathbb{N}, \; t_n- \tau < 0$. Otherwise, when $t_n-\tau \geq 0$,
\[
\begin{aligned}
\partial{\ell}_0(t_n-\tau) &= p\sqrt{2p}  e^{-p(t-\tau)},\\
\partial{\ell}_1(t_n-\tau) &= p\sqrt{2p}  e^{-p(t-\tau)}(2p(t-\tau)-1)) \\
&\qquad -2p\sqrt{2p}e^{-p(t-\tau)},\\
&\vdots
\end{aligned}
\]
etc. Therefore, the ML estimate of the time-delay $\tau$ can be computed as a root of the equation
\[
\sum_{n=0}^{N-1} \left[\left(z_n - \sum_{j=0}^I u_j {\ell}_j(t_n-\tau)\right)\sum_{k=0}^I u_j  \partial{\ell}_j(t_n-\tau)\right] = 0.
\]

Observe that such an estimator is generally biased, and does not come with any finite-sample guarantees. However, it is expected to posses optimal statistical asymptotic properties: consistency (with probability 1) and asymptotical normality, with asymptotic covariance equal to the asymptotic Cram\'{e}r-Rao lower bound.

Unlike in time domain, the ML estimator cannot be realized exactly using the Laguerre domain model, due to the infinite sum in the likelihood. An approximation can be obtained by truncating  the sum. However, the computation of the ML estimate requires solving a non-convex optimization problem, even in time-domain, when we have an explicit expression for the input signal.

\subsection{The Cram\'{e}r-Rao lower bound}
Using the log-likelihood function, the CRLB on the variance of unbiased estimators of $\tau$ given $D_N$ is evaluated  to\footnote{A version of CRLB for biased estimators can also be derived but omitted here.}
\begin{equation}
\begin{aligned}\label{eq:crlb}
\frac{\lambda}{\sum_{n=0}^{N-1} (\partial_\tau \mu_n(\tau))^2} &=\frac{\lambda}{\sum_{n=0}^{N-1} (\partial_\tau u(n\Delta-\tau))^2} \\
&= \frac{\lambda}{\sum_{n=\lfloor \frac{\tau}{\Delta}\rfloor
}^{\lceil\frac{T_u}{\Delta} \rceil } (\partial_\tau u(n\Delta-\tau))^2},
\end{aligned}
\end{equation}
where $\lceil\cdot\rceil, \lfloor\cdot\rfloor$ mean rounding to the nearest integer above and below, respectively.

\subsection{Estimation via frequency interpolation}\label{sec:frequency_interpolation}
Time-delay estimation from discrete (sampled) data is usually performed by maximizing the cross-correlation function between $z_n$ and $u(t_n)$:
\[
\begin{aligned}
r(k) &\triangleq \sum_{n=0}^{N-1} z_{n-k} u(t_n),\\
k^\star &\triangleq \arg\max_{k\in\{0,1,\dots, N-1\}} \;  r(k).
\end{aligned}
\]
Therefore,  the resulting time-delay estimate is  an integer multiple of the sampling time, i.e. $k^\star \Delta$. To estimate a subsample delay, the cross-correlation function $r$ is interpolated in time or frequency domain \citep{Cespedes1995}. Assuming that the initial delay  $k^\star \Delta$ is removed from $r$ by phase shifting, and letting $R_m$ be the discrete Fourier transform obtained after this removal, the subsample delay is evaluated as $k^\star \Delta + \delta\tau$, where $\delta\tau$ satisfies $\text{arg}(R_m) = \omega_m \delta\tau$  and can be estimated by weighted averaging over multiple frequencies. This can be done, e.g., by minimizing  a power-spectrum weighted 2-norm $\sum_{k=1}^{N} w_k( \delta\tau - (\text{arg}(R_k)/\omega_k))^2$  as suggested in \citep[Sec. II, equation~12]{Svilainis2019}. 

\section{Proposed Approach}
\subsection{Estimating the Laguerre spectrum of delay output}
Because $y(t) \in \mathbb{L}_2\lbrack 0, \infty)$, it holds that
\[
y(t) = \sum_{j=0}^\infty y_j \ell_j(t) 
\]
where $\ell_j(t) = \mathcal{L}^{-1}[\ell_j(s)]$ is the inverse Laplace transform (impulse response) of $\ell_j(s)$. The approach developed here is based on an alternative representation of the output signal that does not depend explicitly on $\tau$. For a given  fixed positive integer $K$, we can decompose the signal as follows
\begin{equation}
\begin{aligned}
y(t) &=   \sum_{k=0}^K y_k \ell_k(t) + \tilde{y}(t).
\end{aligned}
\end{equation}
Substituting for $y(t_n)$ in \eqref{eq:measurement_model} yields
\begin{equation}\label{eq:signal_model}
z_n = \sum_{k=0}^K y_k \ell_k(t_n) + \tilde{y}(t_n) + e_n,
\end{equation}
for $n = 0, \dots, N-1$. The observation process can also be expressed using  model \eqref{eq:poly_pure_delay}  that explicitly depends    on $\tau$
\begin{equation}\label{eq:signal_model_explicit}
z_n = \sum_{k=0}^K \sum_{m=0}^{k} \left(h_{j-m}(\kappa) u_{m}\right) \ell_k(t_n) + \tilde{y}(t_n) + e_n.
\end{equation}
These two models constitute the starting point of the approach. Signal model  \eqref{eq:signal_model} can be written in vector form as
\[
Z = \Phi Y + \tilde{\Phi} \beta + E,
\]
where
\[
Z = \begin{bmatrix} z_0\\
z_1\\
\vdots\\
z_{N-1}
\end{bmatrix}, \quad
Y = \begin{bmatrix} y_0\\
y_1\\
\vdots\\
y_{N-1}
\end{bmatrix}, \quad
E = \begin{bmatrix} e_0\\
e_1\\
\vdots\\
e_{N-1}
\end{bmatrix},
\]
\[
\Phi = \begin{bmatrix} \ell_0(t_0) &  \ell_1(t_0)  & \dots  &  \ell_K(t_0)\\
\ell_0(t_1) & \ell_1(t_1)  &   & \ell_K(t_1)\\
\vdots & & \ddots &\\
\ell_0(t_{N-1}) & \ell_1(t_{N-1})& \dots  & \ell_K(t_{N-1})
\end{bmatrix},
\]
and, similarly, for $\tilde{\Phi}$ that is a matrix with infinite number of columns, and $\beta$  that is an infinite-dimensional vector  whose entries represent the tail of the spectrum. In order to construct an estimator of $Y$, we use the following assumption.

{\bf Assumption: } $K$ is sufficiently large, and $\Delta$ is sufficiently small so that $\tilde{y}(t_n) \approx 0, \;\; \forall n$.
\medskip

An estimate of the first part of the output spectrum $Y$ can then be obtained via linear least-squares as
\[
\hat{Y} = \arg\min_Y \; \; \|Z - \Phi Y   \|_2^2.
\]
Assuming that $\Phi$  is well-conditioned (can be checked prior to signal measurement), the estimator is given by the closed-form expression $\hat{Y} = (\Phi^\top \Phi)^{-1} \Phi^\top Z$. Due to the sampling process and the truncation in the Laguerre sum, this estimator will have a small but non-zero bias. If the spectrum of $y$ consists of only $K$ non-zero values, no truncation is needed and then there is no bias.

\begin{prop}
Assume that $u(t) \in \mathbb{L}_2\lbrack 0, \infty)$, $\mathbb{E}[e_k] = 0$  $\forall k$, and  the sampling time $\Delta>0$ as well as the measurement time interval $(N-1)\Delta$ are such that $\Phi$ is full rank. Then, if $\beta \neq 0$, the estimator $\hat{Y}$ of $Y$ is biased.
\end{prop}

\begin{pf}
Notice that $u(t) \in \mathbb{L}_2\lbrack 0, \infty)$ implies $y(t) \in \mathbb{L}_2\lbrack 0, \infty)$.
From the expression for $\hat{Y}$, it follows
\[
\begin{aligned}
\mathbb{E}[\hat{Y}] &=\mathbb{E}[ (\Phi^\top \Phi)^{-1} \Phi^\top (\Phi Y + \tilde{\Phi} \beta + \beta  E)]\\
&  = Y +  (\Phi^\top \Phi)^{-1} \Phi^\top\tilde{\Phi} \beta  + \mathbb{E} [E].
\end{aligned}
\]
Now observe that $\Phi^\top\tilde{\Phi} \neq 0$, but $\mathbb{E}[E] = 0$. The bias is thus given by $(\Phi^\top \Phi)^{-1} \Phi^\top\tilde{\Phi} \beta$ and its value depends on the signal (spectrum of $u$ and $\tau$), the used Laguerre parameter $p$, the sampling time $\Delta$. As $\Delta\to 0$, and taking into account the orthonormality of the Laguerre functions,  the bias becomes ``small".
\end{pf}

The sampled Laguerre functions used to construct $\Phi$ can be easily computed using an impulse invariant transform 
of a continuous-time LTI state-space model of order $K+1$,  whose state $x_j(t) = \ell_j(t)$. Using the definition of the Laguerre function in \eqref{eq:basis}, this LTI model is found to have the state-space matrices
\[
A_c = -2p\begin{bmatrix} 0.5 & 0 & 0 & \dots & 0\\
1 & 0.5 & 0  & \dots & 0\\
1 & 1 & 0.5 & \dots & 0\\
\vdots & & & \ddots \\
1 & 1 & 1 & \dots &0.5 
\end{bmatrix},  B_c = \sqrt{2p} \begin{bmatrix} 1\\
1\\1 \\\vdots \\ 1\end{bmatrix},
\]
and 
$C_c \!= \!\begin{bmatrix} 1 & 1 & \dots & 1\end{bmatrix}$. A discrete-time state-space model whose impulse response matches the sought values at the sampling times is obtained by the matrices
\[
A_d = e^{A_c\Delta}, \;\; B_d = e^{A_c\Delta}B, \;\; C_d  = C_c,\; \text{ and } D_d = B_c.
\]
The $k$-th row of $\Phi^\top$  is then given by
\[
[\Phi]_{k:}^\top = \begin{cases}
D_d,\quad & k = 0\\
C_d A_d^{k-1} B_d,  &0 < k \leq K
\end{cases}
\]

\subsection{Delay estimation in Laguerre domain}
Given the relation between the Markov parameters in \eqref{eq:poly_pure_delay}  and the associated Laguerre polynomials \eqref{eq:assoc_Laguerre}, it is possible to obtain a closed-form expression for $\tau$ in terms of any three consecutive Markov parameters. The expressions for the continuous-time case  are similar in structure to those derived for the discrete-time case in \cite{Abdalmoaty2022}.

Solving \eqref{eq:three-term-cont}, when $\xi = \kappa$, for $\kappa$ and substituting the Markov parameters $h_m(\kappa) = e^{-\frac{\kappa}{2}} {\rm L}_m(\kappa)$ we obtain the expression
\[
\tau = -\frac{(m+1) h_{m+1} + (m-1) h_{m-1} - 2mh_m}{h_m}
\]
for any $m\geq 1$. This relation then leads to
\[
\begin{aligned}
\kappa h_0 & = - h_1,\\
\kappa h_1 & = -2 h_2 + 2 h_1,\\
\kappa h_2 & = -3 h_3 + 4 h_2 - h_1,\\
\kappa h_3 & = -4 h_4 + 6 h_3 -2 h_2,\\
&\vdots\\
\kappa h_{M\!-\!2} & = -(M\!-\!1) h_{M\!-\!1} + 2(M\!-\!2) h_{M\!-\!2} - (M\!-\!3) h_{M\!-\!3}.
\end{aligned}
\]

Suppose that the first $M$ Markov parameters are available, namely $h_0, \dots, h_{M-1}$. Define the following two  vectors
\[
A \triangleq \Omega \begin{bmatrix}
h_0\\
\vdots\\
h_{M-2}
\end{bmatrix}\! -\! (M-1) \begin{bmatrix}
0\\
\vdots\\
0\\
h_{M-1}
\end{bmatrix}, \qquad B \triangleq  \begin{bmatrix}
h_{0}\\
\vdots\\
h_{M-2}
\end{bmatrix},
\]
and the tridiagonal matrix $\Omega \in \mathbb{R}^{{M-1}\times {M-1}}$ 
\[
\Omega \triangleq \begin{bmatrix}
0 & -1 & 0  & 0& \dots & 0 &0 \\
0 & 2 & -2 & 0 &  \dots & 0& 0\\
0 & -1 & 4 & -3 & \hdots & 0&0 \\
0 & 0 & -2 & 6 & \hdots & 0&0 \\
\vdots &  &  & & \ddots & \vdots & \vdots\\
0 &  0&  0 & 0& \hdots & 2(M-2) & -(M-1)\\
0 & 0 & 0 & 0 &\hdots& -(M-3) & 2(M-1)\\
\end{bmatrix}.
\]
Then, the time-delay satisfies the vector equation $A = 2p B\, \tau$
and can be computed as
\begin{equation}\label{eq:delay_noisefree}
\tau = \frac{1}{2p}\frac{B^\top A}{B^\top B}.
\end{equation}
To evaluate  $\tau$ using this formula, we need to estimate  the vectors $A$ and $B$. Since the Laguerre spectrum of $u(t)$ is known, and an estimate of the Laguerre spectrum of the output signal is available, we can devise a linear least-squares estimate of the Markov parameters as follows.

From model  \eqref{eq:poly_pure_delay}, we have
\[
Y = T(U)H,
\]
\[
Y\! =\! \begin{bmatrix} y_0\\ y_1\\ \vdots\\ y_K\end{bmatrix}, \; 
T(U) \!=\! 
\begin{bmatrix} u_0 & 0 & \dots & 0\\
u_1 & u_0 & \dots & 0\\
\vdots & \vdots & \ddots\\
 u_K & u_{K-1} & \dots & u_0\end{bmatrix}, \; 
H\! =\! \begin{bmatrix} h_0\\ h_1\\ \vdots\\ h_K\end{bmatrix},
\]
and $T(U)$ is non-singular whenever $u_0 \neq 0$. While $U$ is known, the spectrum $Y$ is unknown but is estimated; we replace the left hand side by its estimate plus the bias and the random error
\[
\overbrace{\hat{Y} + \underbrace{(Y-\mathbb{E}[\hat{Y}])}_{\text{bias}} - \underbrace{\mathcal{E}}_{\text{zero mean  error}} }^{{Y}} = T(U)H.
\]
Neglecting the bias term and rewriting the expression gives
\[
\hat{Y} = T(U)H +\mathcal{E}.
\]
Then the least-squares estimate of $H$ is
\[
\hat{H} = T^{-1}(U)\hat{Y}
\]
This estimate is unbiased whenever $\hat{Y}$ is, and it can be used to estimate $A$ and $B$, and construct an estimator of the delay.

\subsection{Accuracy analysis of the  estimators}
To summarize the exposition in the previous section; The models in time and Laguerre domain lead to the following relations:
\begin{equation}
\begin{aligned}
Z &= \Phi Y + \tilde{\Phi} \beta + E,\\
Y &= T(U)H,\\
\tau &= \frac{1}{2p}\frac{B^\top A}{B^\top B},
\end{aligned}
\end{equation}
where the known values are $\Phi$, $T(U)$, $p$, and $A$ and $B$ as functions of the unknown $H$. These relations are then ``inverted" to obtain $\hat{\tau}$ as follows
\begin{equation}
\begin{aligned}\label{eq:delay_estimator}
\hat{H} &= T^{-1}(U) (\Phi^\top \Phi)^{-1} \Phi^\top Z,\\
\hat{\tau}& =  \frac{1}{2p}\frac{B(\hat{H})^\top A(\hat{H})}{B(\hat{H})^\top B(\hat{H})}.
\end{aligned}
\end{equation}
In particular, 
\[
\begin{aligned}
\hat{H} &= H + \tilde{H}\\
&= H + T^{-1}(U) (\Phi^\top \Phi)^{-1} \tilde{\Phi}^\top \beta + T^{-1}(U) (\Phi^\top \Phi)^{-1} \Phi^\top E,
\end{aligned}
\]
where $\tilde{H}$ denotes the errors in $\hat{H}$. The second term of the last equation represents the bias (deterministic error) which depends on the true unknown delay via $\beta$, and the third term gives the random error. The covariance of $\hat{H}$ is given by
\[ \mathrm{cov} \hat{H}=
\lambda T^{-1}(U) (\Phi^\top \Phi)^{-1}T^{-\top}(U).
\]
The mean-square error is thus
\begin{equation}\label{eq:mse}
\begin{aligned}
\text{MSE}(\hat{H}) &= \|T^{-1}(U) (\Phi^\top \Phi)^{-1} \tilde{\Phi}^\top \beta \|_2^2 \\
& \;\; + \lambda\; \text{tr}\,( T^{-1}(U) (\Phi^\top \Phi)^{-1}T^{-\top}(U)).
\end{aligned}
\end{equation}

The estimator of $\tau$ in \eqref{eq:delay_estimator} is thus biased even when  $\hat{H}$ is unbiased. We now clarify this by showing that we have an \textit{errors-in-variables} case (see \cite{Soderstrom2018}), and work out an expression for the bias.

Define $\vec{1}_{M-1} \triangleq
\begin{bmatrix}0 &0& \dots& 1 \end{bmatrix}^\top
\in \mathbb{R}^{M-1}$
\[
\hat{B} =  B(\hat{H})  = [\hat{H}]_{1:M-1} = \begin{bmatrix}\hat{h}_0\\ \hat{h}_1\\ \vdots\\ \hat{h}_{M-2}   \end{bmatrix} = B +  
\overbrace{\begin{bmatrix}\tilde{h}_0\\ \tilde{h}_1\\ \vdots\\ \tilde{h}_{M-2} \end{bmatrix}}^{E_B},  
\]
\[
\begin{aligned}
\hat{A} =  A(\hat{H})  &= \Omega \hat{B} - (M-1)\tilde{h}_{M-1} \vec{1}_{M-1} \\
& = A + \underbrace{\Omega E_B - (M-1) \tilde{h}_{M-1}\vec{1}_{M-1} }_{E_A}.
\end{aligned}
\]
Thus, one can now see that we have the following errors-in-variables model
\[
\begin{aligned}
A &=  2pB\; \tau,\\
\hat{A} &= A + E_A,\\
\hat{B} &= B + E_B,
\end{aligned}
\]
where  the ``measurements" are $\hat{A}$ and $\hat{B}$, and the noises $E_A$ and $E_B$ are correlated.
\medskip

\begin{thm}
     The bias of the time-delay estimator $\hat{\tau}$ in \eqref{eq:delay_estimator}  is given by
    \[
   \frac{1}{2p} \mathbb{E}\left[\frac{\varepsilon_1}{B^\top B +\varepsilon_2}\right] - \tau\;\mathbb{E}\left[\frac{\varepsilon_2}{B^\top B + \varepsilon_2} \right],
    \]
    where
    \[
\begin{aligned}
\varepsilon_1 &= E_B^\top A + E_A^\top B + E_B^\top E_B,\\
\varepsilon_2 &= 2 E_B^\top B + E_B^\top E_B.\\
\end{aligned}
    \]
\end{thm}
\medskip
\begin{pf}
The expression is obtained using straightforward algebraic manipulations. Notice that
\[
\begin{aligned}
2p \hat{\tau}& =  \frac{\hat{B}^\top \hat{A}}{\hat{B}^\top \hat{B}} = \frac{(B+E_B)^\top(A+E_A)}{(B+E_B)^\top(B+E_B)}\\
&= \frac{B^\top A+ E_B^\top A + E_A^\top B + E_B^\top E_B}{B^\top B+ 2 E_B^\top B + E_B^\top E_B} = \frac{B^\top A+ \varepsilon_1}{B^\top B+ \varepsilon_2}.
\end{aligned}
\]
Therefore,
\[
\begin{aligned}
 \hat{\tau}& = \tau + \frac{1}{2p}\frac{B^\top A+ \varepsilon_1}{B^\top B+ \varepsilon_2} -\tau = \tau + \frac{1}{2p}\left(\frac{B^\top A+ \varepsilon_1}{B^\top B+ \varepsilon_2} -2p\tau\right)\\
 & = \tau + \frac{1}{2p}\left(\frac{B^\top A+ \varepsilon_1 - B^\top A - 2p\tau \varepsilon_2}{B^\top B+ \varepsilon_2} \right).\\
\end{aligned}
\]
Then,
\[
\text{Bias}(\hat{\tau}) = \mathbb{E}[\hat{\tau}] - \tau = \frac{1}{2p}\mathbb{E}\left[\frac{ \varepsilon_1- 2p\tau \varepsilon_2}{B^\top B+ \varepsilon_2} \right].
\]
\end{pf}
\medskip

As expected, the bias depends on the unknown true delay value in a non-trivial way via the true Markov parameters. Notice that $B$ depends on both $\tau$ and  $p$, and thus the dependence of the bias on $p$ comes in a non-trivial form as well. Yet, the expression is useful and could potentially be used for bias correction. When the measurement noise is Gaussian, the expectations in the bias expression will be with respect to a non-central correlated normal ratio density (which has a known form, as shown in \cite[p. 636]{hinkley1969}). Thus, for a given $p$ and $\tau$, the expectations can be approximated using numerical integration. An alternative is to use a Monte-Carlo approximation. Then, given $p$, $\lambda$ and $\check{\tau}$, the bias can be predicted.

\subsection{Experiment design}\label{sec:experiment_design}
In order to pose a well-defined problem, both the spectrum of the input and the Laguerre parameter $p$ need to be tuned a priori to the chosen experimental settings. This can be done once the sampling time $\Delta$ and the number of samples $N-1$ are decided. Then, we assume that the number of Laguerre functions used for signal approximation is fixed to some finite positive integer, and that a rough estimate $\check{\tau}$ of the time-delay is available.

Since the final objective is to estimate the delay $\tau$, an ideal experiment design would minimize the MSE of $\hat{\tau}$. 
However, computing the MSE of $\hat{\tau}$ in \eqref{eq:delay_estimator} is not an easy task. 
Here, we will be content with maximizing the accuracy of the Markov parameters  estimator $\hat{H}$ (minimize the MSE of $\hat{H}$), and save the study of more advanced designs for a future contribution.  Notice that if the measurement noise is Gaussian, $\hat{H}$ will coincide  with the maximum-likelihood estimator of $H$ only if its bias is zero. But then, by invoking the invariance property of the maximum-likelihood estimator, the efficiency is preserved asymptotically; namely since $\hat{H}$ converges to the true ones, the estimate $\hat{\tau}$ is expected to converge to $\tau$ as well. Based on these  arguments, we propose the following experiment design
\[
\begin{aligned}
\left(p^\star, \{u_k^\star\}_{k=1}^I\right) =  \;\;
& \arg\min_{p, u_0, \dots, u_I}
& & \text{MSE}\left(\hat{H}(p, \{u_k\}_{k=1}^I) \;|\; \check{\tau} \right)\\
& \text{subject to}
& & p>0,\quad u_0 > 0,\\
& & &  u_k \geq 0 \quad \quad \;\, \text{ for odd } k,\\
& & & u_k = -u_{k-1} \text{ for even } k,\\
& & & \sum_{k=0}^I u_k^2 \leq \eta,
\end{aligned}
\]
for an odd value $I$, and a given upper bound $\eta$ on the input energy and initial estimate/guess $\check{\tau}$. Notice that $\sum_{k=0}^I u_k^2$ is the squared 2-norm of $u(t)$ by Parseval's  theorem.

The MSE in the objective functions is that of $\hat{H}$ defined in \eqref{eq:mse}. The variance term in the MSE is easily evaluated using the discrete-time state-space model obtained by  impulse invariant transform. The bias term, however, cannot be evaluated using the expression given by \eqref{eq:mse} (recall that $\beta$ is infinite dimensional). Nevertheless, for a given $\check{\tau}$, the bias can be evaluated by simulating a noise-free output trajectory $y(t)$, finding its projection $\hat{Y}$, computing $T^{-1}(U)\hat{Y}$, and subtracting it from the vector $H$ constructed using the associated Laguerre polynomials in \eqref{eq:assoc_Laguerre}. 

A remark on the constraints on $\{u_k\}$ is in place here: The proposed constraints ensure that $u(0) = 0$, and thus the continuity of the output signal. The reason for this is to avoid the Gibbs phenomenon and reduce the number of parameters used to tune the input. 

We would like to stress here that the above experiment design problem is to be solved offline; iterative numerical solvers can be employed. Alternatively, when $I$ is a small number (say 3 -- then total number of unknowns = 3), the problem can be handled efficiently by gridding. 

\section{The Algorithm}
A step-by-step summary of the proposed algorithm is provided in this section. Given
\begin{itemize}
\item sampling time $\Delta$, and measurement interval $[0,T]$
\item initial estimate of the delay (e.g., a multiple of $\Delta$)
\item number of Laguerre functions used to design $u(t)$, denoted $I+1$
\item constraint on the energy of $u(t)$ over $[0,T]$
\item number of Laguerre functions used to construct the signal model of $y(t)$, denoted $K$. Usually $K\geq I$,
\end{itemize}
\smallskip
the algorithm is executed as follows:\smallskip
\begin{enumerate}
\item Experiment design (offline): tune the Laguerre parameter $p$, and the continuous Laguerre spectrum of the input signal $\{u_j\}_{j=0}^I$  by minimizing the MSE in \eqref{eq:mse}, subject to the given constraints on the energy of the input signal and such that $u_1>0$, $u_k \geq 0$ for odd $k$ and, $u_m = -u_{m-1}$ for even $m \in \mathbb{N}_+$.
\item Construct $u(t), \; 0\leq t \leq T$, by realizing the Laguerre filters, then apply it to the system, measure the output, and construct $Z$.
\item Compute $\hat{H}$ and $\hat{\tau}$ according to \eqref{eq:delay_estimator}
\end{enumerate}

\section{Numerical Experiments}

\subsection{Noise-free Simulations}
To study the bias, we ran noise-free simulations with the following settings:
\begin{itemize}
\item time resolution for continuous time simulation  1 $\mu s$,
\item SNR = 46.0205 dB: noise variance = 0.01, with constraints on the input spectrum as described above with  $I = 3$ and $\eta = 2$,
\item $T=0.5 s$ and $K$ = 6,
\end{itemize}
and solved the optimal design for each of the following sampling times
\[
\begin{aligned}
\Delta \in \{1\!\times\!10^{-4},\;\; 8\!\times\!10^{-5}\;\;, 6\!\times\!10^{-5} \}
\end{aligned}
\]
using the initial guess $\check{\tau} = \Delta$. The considered true time-delays are
\[
\begin{aligned}
\tau &\in \{3\!\times\!10^{-5},\;\;  2\!\times\!10^{-5},\;\;  1\!\times\!10^{-5}\}.
\end{aligned}
\]
All values are in seconds. Thus, we have the case where $\check{\tau}$ overestimates $\tau$, which lies within the first sampling interval. Then we evaluate the biases in $\hat{H}$ and $\hat{\tau}$. 

\begin{figure}
    \centering
    \includegraphics[width=0.37\textwidth]{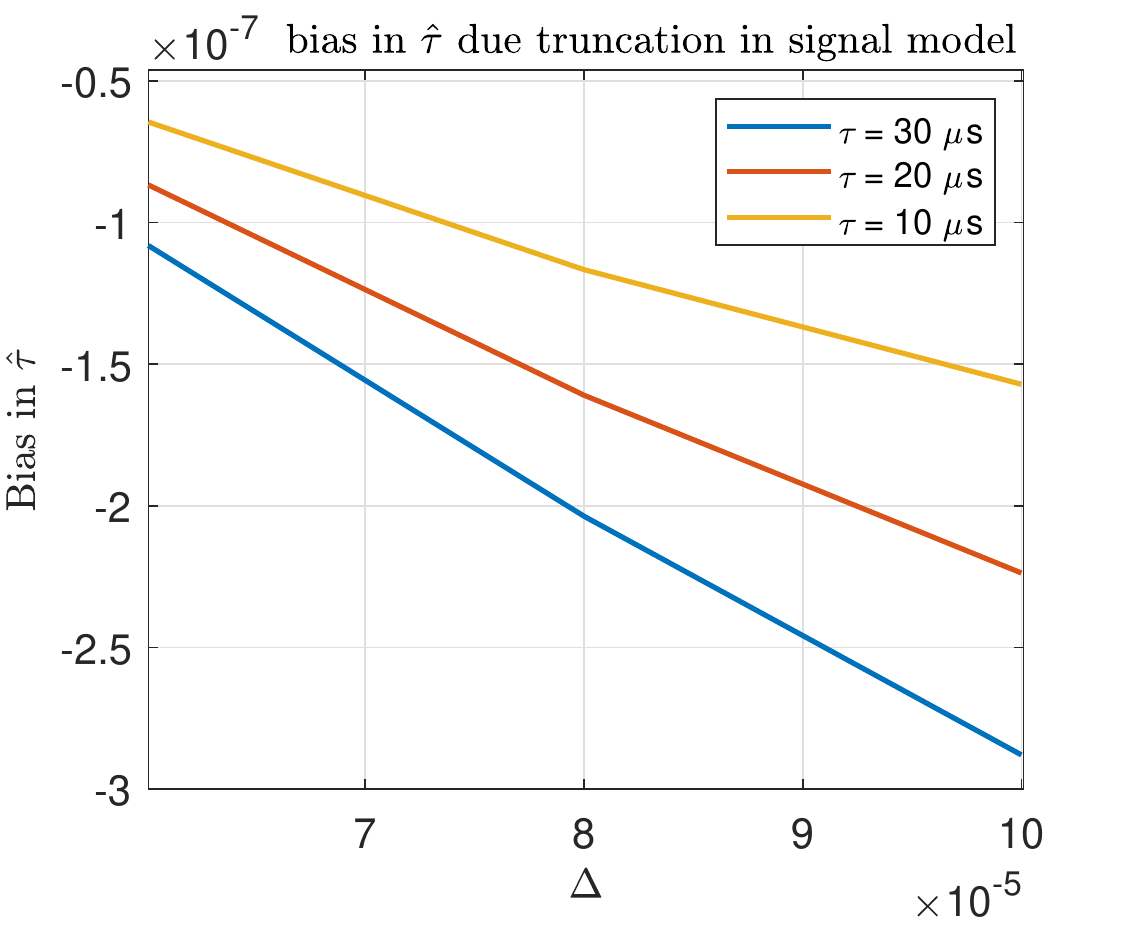}
    \caption{Bias in $\hat{\tau}$ due to the truncation of the continuous Laguerre spectrum of the output at the finite value $K = 6$. A linear growth of the bias is observed with growing sampling time $\Delta$.}
    \label{fig:biasInTau}
\end{figure}

\begin{figure}
    \centering
    \includegraphics[width=0.37\textwidth]{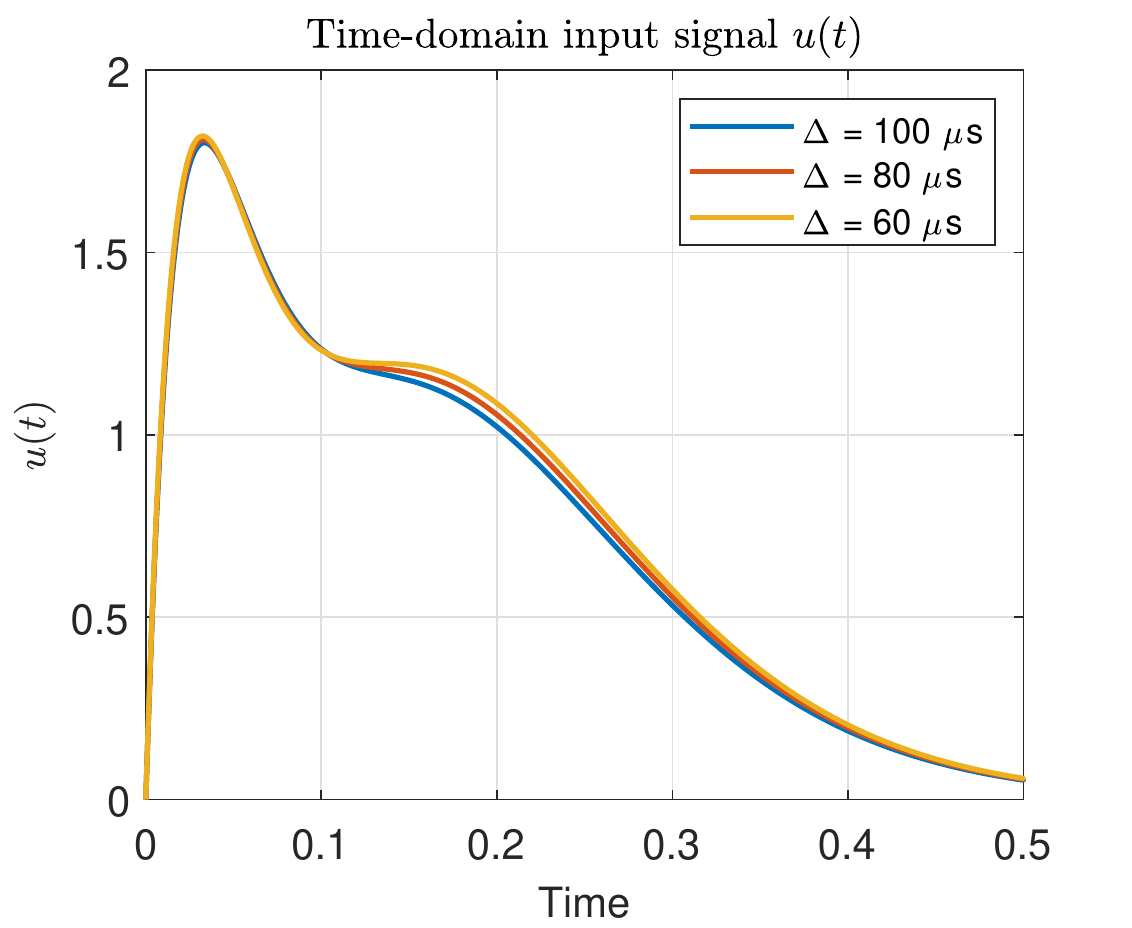}
    \caption{Time-domain  input  for three values of $\Delta$. The corresponding values of the Laguerre parameter are all around $p =  20$. Those were found by using local optimization routines to get an approximate (local) solution for the problem in Section~\ref{sec:experiment_design}}
    \label{fig:optimalInput}
\end{figure}

\begin{figure}
    \centering
    \includegraphics[width=0.37\textwidth]{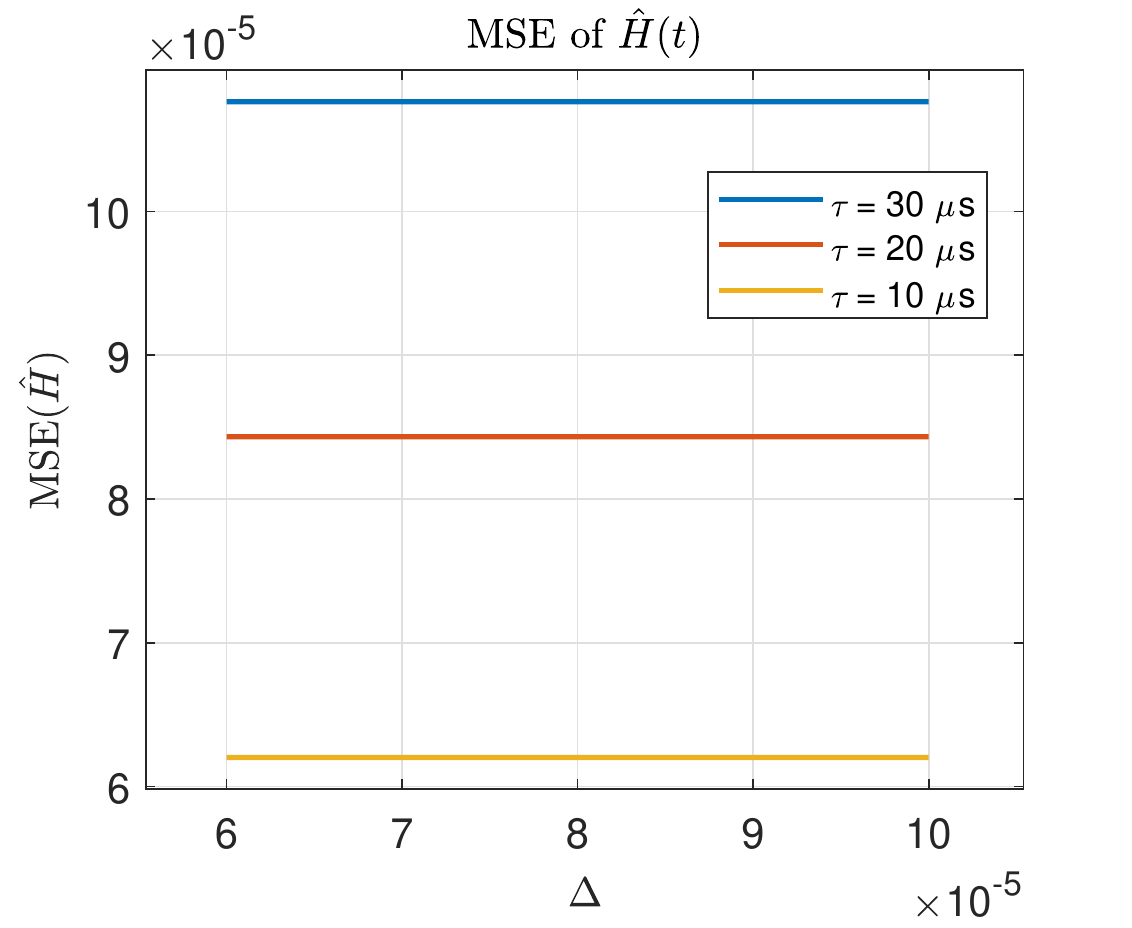}
    \caption{MSE of the Markov parameter estimator. Notice that for fixed $\tau$, the bias increases with $\Delta$. The change is small and not evident due to the used y-axis scale.   }
    \label{fig:MSEHhat}
\end{figure}

The obtained results are presented in Fig.~\ref{fig:biasInTau}, Fig.~\ref{fig:optimalInput}, and Fig.~\ref{fig:MSEHhat}. As expected, the bias in $\hat{\tau}$ decreases with decreasing sampling time, and smaller bias is achieved for the more accurate initial delay estimates, as shown in Fig.~\ref{fig:biasInTau}. Similar observations hold for the MSE of $\hat{H}$ as shown in Fig.~\ref{fig:MSEHhat}. The corresponding time-domain input signals are shown in Fig.~\ref{fig:optimalInput}, for the three considered sampling time values.

\subsection{Monte-Carlo Simulations}
\begin{figure}
    \centering
    \includegraphics[width=0.35\textwidth]{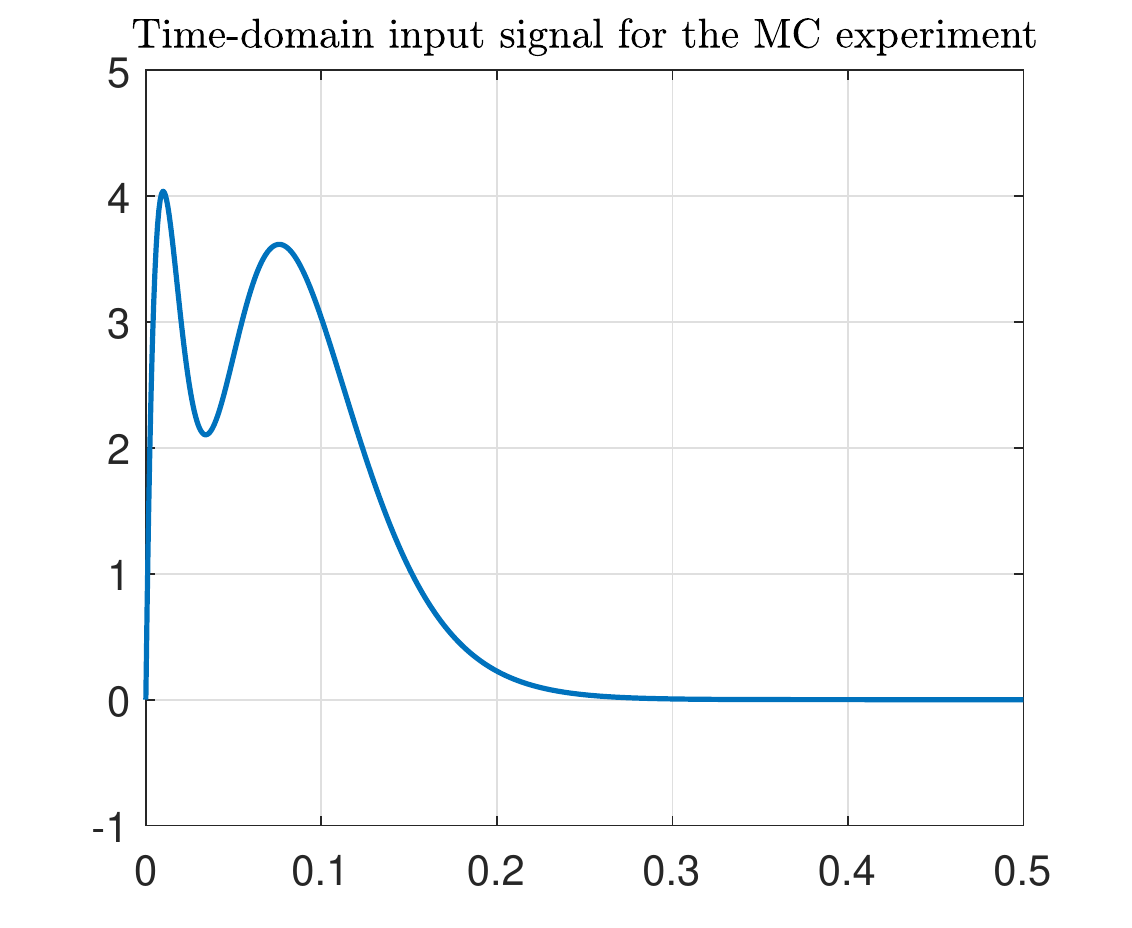}
    \caption{Time-domain input signal $u(t)$ used in the Monte-Carlo simulation study}
    \label{fig:time_domain_input}
\end{figure}

We also ran a Monte-Carlo simulation study where we compared the performance of the considered algorithm to i) the maximum-likelihood estimator derived in Section~\ref{sec:MLE}, ii) Laguerre-domain $\hat{\tau}$ estimator where $\{y_j\}$ are estimated by approximating the defining integral \eqref{eq:spectrum_definition} using cubic splines, iii) Frequency-domain interpolation method as described in Section \ref{sec:frequency_interpolation} - we used the best reported method in \cite{Svilainis2019} (weighted $L_2$ norm), and the MATLAB code given there. Our experiment setup is as follows: Time resolution for continuous-time simulation = $10\mu s$, true time-delay $\tau = 0.00133$, $T = 0.5 s$, $\Delta  = 0.0003 s$ and the discrete-time noise variance is 0.01. The number of Laguerre functions used to design the input is 4  ($I = 3$), and $\eta = 2$ (SNR = 200). The initial guess of the time-delay in the experiment design $\check{\tau} = \Delta$ (which is quite far from the true value). The used signal $u(t)$ is shown in Fig~.\ref{fig:time_domain_input} where the Laguerre parameter $p\approx 50$.   Finally, the number of Laguerre functions used for the output signal model is set to 11 ($K = 12$).

\begin{figure}
    \centering
    \includegraphics[width=0.5\textwidth]{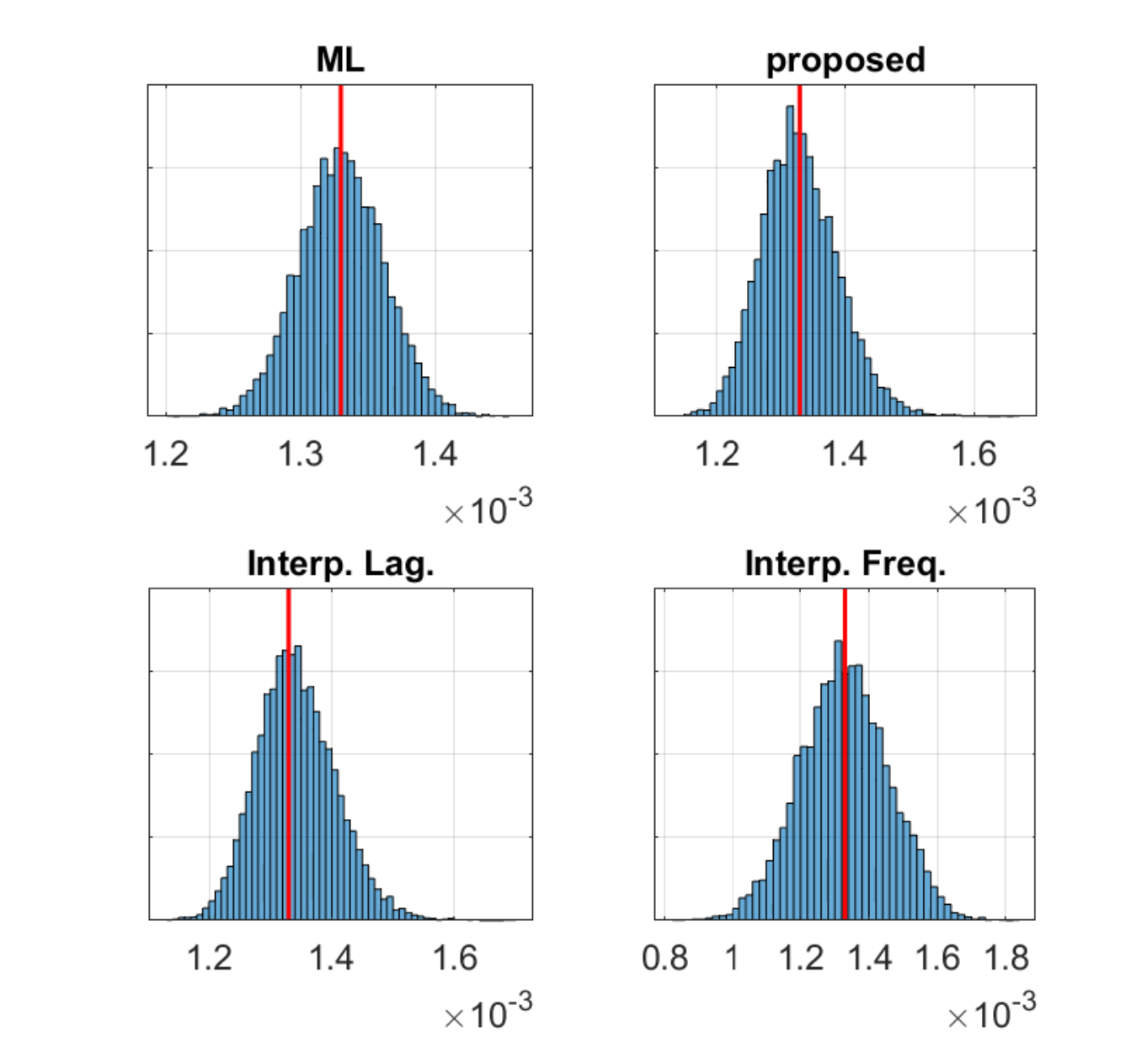}
    \caption{Histograms of the four considered estimators. The red line indicates the true value of the delay.  }
    \label{fig:hist}
\end{figure}

The experiment is performed over 10000 independent data sets; the results in terms of the bias, variance and (normalized) MSE are summarized in Table \ref{tab:monte_carlo_results}. We also included the histograms of the four estimators in Fig.~\ref{fig:hist}.

\begin{table}[!ht]  
  \centering
  \begin{tabular}{cccc}
    \toprule
                     & Bias                         & Var                          & MSE                   \\ \midrule
    \tiny{(discrete-time)}\\ CRLB; see \eqref{eq:crlb} \vspace{-0.0cm}  \vspace{0.2cm} &  &                              & $1.011\!\times\!10^{-9}$  \\ 
    ML               & $4.016\!\times\!10^{-7}$     & $1.001\!\times\!10^{-9}$     & $4.090\!\times\!10^{-8}$   \\
    Proposed         & $5.507\!\times\!10^{-8}$     & $3.592\!\times\!10^{-9}$      & $1.466\!\times\!10^{-7}$   \\ 
    Interp. Lag.     &  $1.306\!\times\!10^{-5}$    & $4.121\!\times\!10^{-9}$     & $1.752\!\times\!10^{-7}$  \\ 
    Interp. Freq.    & $-1.985\!\times\!10^{-6}$    & $1.660\!\times\!10^{-8}$     & $6.781\!\times\!10^{-7}$   \\ 
    \bottomrule
  \end{tabular}
    \caption{Bias, variance and normalized MSE (by a factor $\sqrt{N}$) of $\hat{\tau}$ for the different  methods. }
  \label{tab:monte_carlo_results}
\end{table}

\section{Conclusion}
An approach to the estimation of the pure continuous time delay from sampled output measurements is proposed. It is based on the Laguerre-domain relationship between the spectra of the input and output and consists of two consecutive steps: First the continuous Laguerre spectrum of the output is estimated from sampled noisy measurements. Then the delay value is estimated from the spectra of the input and the output using the concept of Markov parameters of the delay operator. Although the developed approach demands elaborate tuning, it is shown in a numerical simulation example  to exhibit  high performance in comparison with state-of-the-art continuous delay estimation algorithms, for the particular kind of signals used.
~~ 

\bibliography{bibliography}          
\end{document}